\documentclass[conference,10pt]{IEEEtran}
\usepackage{epsfig,rotating,setspace,latexsym,amsmath,epsf,amssymb,amsfonts,bm,theorem,subfigure,epstopdf}
\usepackage{cite,authblk}
\usepackage{bbm}
\usepackage{algorithmic,algorithm}
\newtheorem{theorem}{Theorem}
\newtheorem{lemma}{Lemma}
\newenvironment{Proof}[1]{\medskip\par\noindent{\bf Proof:\,}\,#1}{{\mbox{\,$\blacksquare$}\par}}
\IEEEoverridecommandlockouts
\newtheorem{corollary}{Corollary}
\allowdisplaybreaks
\usepackage{color}
\usepackage{graphicx}
\usepackage[utf8]{inputenc}
\usepackage{amsmath}
\usepackage{calc}

\begin{document}
	
	\title{Age of Information in Multicast Networks with Multiple Update Streams \thanks{This work was supported by NSF Grants CNS 15-26608, CCF 17-13977 and ECCS 18-07348.}}
	
	\author[1]{Baturalp Buyukates}
	\author[2]{Alkan Soysal}
	\author[1]{Sennur Ulukus}
	\affil[1]{\normalsize Department of Electrical and Computer Engineering, University of Maryland, MD}
	\affil[2]{\normalsize Department of Electrical and Electronics Engineering, Bahcesehir University, Istanbul, Turkey}
	
	\maketitle
	
\begin{abstract}	
We consider the age of information in a multicast network where there is a single source node that sends time-sensitive updates to $n$ receiver nodes. Each status update is one of two kinds: type I or type II. To study the age of information experienced by the receiver nodes for both types of updates, we consider two cases: update streams are generated by the source node at-will and update streams arrive exogenously to the source node. We show that using an earliest $k_1$ and $k_2$ transmission scheme for type I and type II updates, respectively, the age of information of both update streams at the receiver nodes can be made a constant independent of $n$. In particular, the source node transmits each type I update packet to the earliest $k_1$ and each type II update packet to the earliest $k_2$ of $n$ receiver nodes. We determine the optimum $k_1$ and $k_2$ stopping thresholds for arbitrary shifted exponential link delays to individually and jointly minimize the average age of both update streams and characterize the pareto optimal curve for the two ages.
\end{abstract}

\section{Introduction}
Age of information is a metric that measures the freshness of the received information. A typical model to study age of information includes a source which acquires time-stamped status updates from a physical phenomenon. These updates are transmitted over a network to the receiver(s) and age of information in this network or simply the \emph{age} is the time elapsed since the most recent update at the receiver was generated at the transmitter. In other words, at time $t$, age $\Delta(t)$ of a packet which was generated at time $u(t)$ is $\Delta(t) = t-u(t)$. Age of information has been extensively studied in a queueing-theoretic setting in references \cite{Kaul12b, Costa14, Bedewy16, Sun17a, Najm17, Kadota16} and in an energy harvesting setting in references \cite{Arafa17b, Arafa17a, Bacinoglu15, Wu18, Arafa18a, Arafa18b, Arafa18d, Baknina18a, Baknina18b}.

\begin{figure}[t]
	\centering  
	\includegraphics[width=.95\columnwidth]{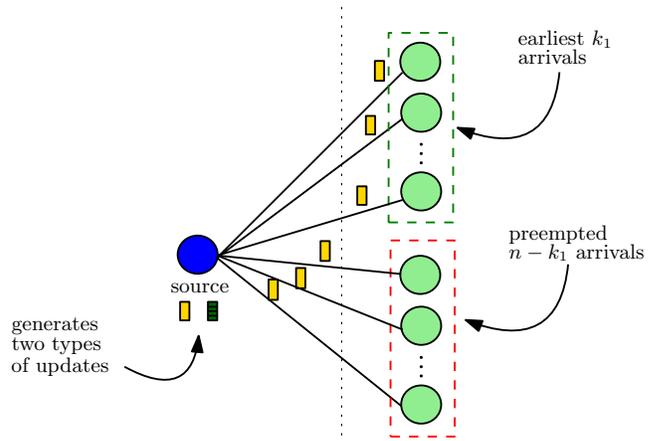}
	\caption{Multicast network with a single source node sending two types of updates to $n$ nodes.}
	\label{fig:model}
	\vspace{-0.5 cm}
\end{figure}

With the increase in the number of users in networks supplying time-sensitive information, the scalability of age as a function of the number of nodes has become an issue. To this end, reference \cite{Zhong17a} studies a single-hop multicast network in which time-sensitive information packets are sent to a large number of interested recipients simultaneously and shows that appropriate stopping threshold $k$ can prevent information staleness as the network grows. References \cite{Buyukates18, Buyukates18b} extend this result to single-hop multicast networks with exogenous updates and to multihop multicast networks and show that by utilizing transmission schemes with stopping thresholds, the age of information at the users can be made a constant as the network grows. In all these works, there is only one type of update that the users are interested in. In a more practical scenario there could be many streams sharing the same network. For example, in an autonomous vehicle network, the network can carry information about the velocity, position and acceleration of a car and broadcast it to all nearby cars. Thus, often networks are used to transmit multiple update streams which include multiple different types of update packets. References \cite{Huang15, Kaul18, Najm18a, Najm18b} study the age of information at a monitor node which receives multiple streams of update packets with or without different priorites. 

In this work, we study the age of information in a multicast network with multiple update streams. Two types of updates are transmitted from the source node to $n$ receiver nodes: type I and type II. We are interested in the average age experienced by the receiver nodes for both of the update streams. We first analyze the setting in which the updates are generated by the source node at-will and show that the average age of either of the update streams at an individual node can be made a constant independent of $n$ using the earliest $k_1$, $k_2$ transmission scheme such that each type I update is transmitted to the earliest $k_1$ of $n$ nodes and each type II update is sent to the earliest $k_2$ of $n$ nodes. We determine the optimal stopping thresholds to individually and jointly optimize the average age of type I and type II updates at the receiver nodes for arbitrary shifted exponential link delays and characterize their pareto optimal curve. Then, we extend these results to the case in which the updates arrive exogenously to the source node. 

\section{System Model and Age Metric} \label{model}
We consider a system (see Fig.~\ref{fig:model}), where there is a single source node broadcasting time-stamped updates to $n$ nodes using $n$ links with i.i.d.~random delays. Each status update is one of two kinds: type I or type II. Thus, in this system two different update streams share the same network. Each of the $n$ receiver nodes is interested in both streams. A type I update takes $X$ time to reach from the source node to a particular receiver node whereas a type II update needs $\tilde{X}$ time to reach to a receiver node where $X$ and $\tilde{X}$ are shifted exponential random variables with parameters $(\lambda,c)$ and $(\tilde{\lambda},\tilde{c})$, respectively, where $c$ and $\tilde{c}$ are positive constants. Different update types have different service rates considering their possibly different lengths, compression rates etc. 

We consider two variations on the operation of this network. In the first case, updates are generated by the source node at-will. At each time the source node generates a type I update with probability $p_1$ or a type II update with probability $p_2$ where $p_1+p_2=1$. In the second case, however, both of the update streams arrive exogenously to the source node. Here, we model the exogenous update arrival as a Poisson process with rate $\mu$ where each arriving update is a type I update with probability $p_1$ or a type II update with probability $p_2$ such that $p_1+p_2=1$. Thus, type I update stream arrives as a Poisson process with rate $\mu_1 = \mu p_1$ and type II stream arrives as a Poisson process with rate $\mu_2 = \mu p_2$.

In either way of operation, the source node adapts the earliest $k_i$ transmission scheme where $i =1,2$ depending on the type of the update. Assume the $j$th update that is sent from the source node is of type I. Then, the source node waits for the acknowledgment from the earliest $k_1$ of $n$ receiver nodes. After it receives all $k_1$ acknowledgement signals, we say that update $j$ has been completed. At this time, transmissions of the remaining $n-k_1$ packets are terminated. If the source node generates the updates itself, it generates the update $j+1$ as soon as the update $j$ has been completed, i.e., the source node implements a zero-wait policy. However, if the updates arrive exogenously, the source node starts to wait for the next update arrival as soon as update $j$ has been completed. When a type II update is transmitted from the source node, this process is repeated with stopping threshold $k_2$ instead of $k_1$.

Since we have two different update streams, each receiver node experiences two different age processes. Thus, age is measured for each of the $n$ receiver nodes for each update type. Age of type I updates at node $i$ at time $t$ is the random process $\Delta_{I,i}(t) = t - u_{I,i}(t)$ where $u_{I,i}(t)$ is the time-stamp of the most recent type I update at that node. The metric we use, time averaged age, is given by
\begin{align}
\Delta_I = \lim_{\tau\to\infty} \frac{1}{\tau} \int_{0}^{\tau} \Delta_I(t) dt
\end{align}
where $\Delta_I(t)$ is the instantaneous age of the last successfully received type I update as defined above. Age of type II updates, $\Delta_{II}$, is defined accordingly. We will use a graphical argument to derive the average age at an individual node for each update type. Since all link delays are i.i.d. for all nodes and packets of the same kind, each node $i$ experiences statistically identical age processes and will have the same average age for either update type. Therefore, it suffices to focus on a single receiver node for the age analysis.

\section{Age Analysis}
\subsection{At-will Update Generation} \label{atwill}
The source node generates updates when the previous update is completed. Independent from previous link delays and updates, each generated update belongs to type I with probability $p_1$ and type II with probability $p_2$ such that $p_1+p_2=1$. As described in Section~\ref{model}, the source node implements the earliest $k_1$ and $k_2$ transmission policy for type I and type II updates, respectively, where $k_1, k_2 \in \{1,2, \dots, n\}$. Thus, in this section we extend the result from \cite{Zhong17a} to multiple update streams using the same multicast network.

We denote the time between any two update departures from the source node as the update cycle and represent it with random variable $Y$. Since the source node adapts a zero-wait policy and generates the next update right after the current update has been completed, update cycle of an update is equal to the transmission time of that update. If the $j$th update is of type I, its update cycle, $Y_j$, is the time needed to reach $k_1$ out of $n$ receiver nodes which is equal to $X_{k_1:n}$. Correspondingly, if the $j$th update is of type II, $Y_j$ is equal to $\tilde{X}_{k_2:n}$. Thus,
\begin{align}
Y_j = \begin{cases} 
X_{k_1:n} & w.p.~p_1 \\
\tilde{X}_{k_2:n} & w.p.~p_2 
\end{cases} \label{trans_time}
\end{align}
We denote the $k$th smallest of $X_1, \dots, X_n$ as $X_{k:n}$. For a shifted exponential random variable $X$, we have
\begin{align}
E[X_{k:n}] =& c + \frac{1}{\lambda}(H_n - H_{n-k}) \label{ord1}\\
Var[X_{k:n}] =& \frac{1}{\lambda^2}(G_{n} - G_{n-k}) \label{ord2}
\end{align}
where $H_n = \sum_{j=1}^{n} \frac{1}{j}$ and $G_{n} = \sum_{j=1}^{n} \frac{1}{j^2}$. Using these,
\begin{align}
E[X_{k:n}^2] =& c^2 + \frac{2c}{\lambda}(H_n - H_{n-k})  \nonumber \\
&+ \frac{1}{\lambda^2}\left((H_n - H_{n-k})^2 +G_{n} - G_{n-k} \right) \label{ord3}
\end{align}

A type I (type II) update that is sent from the source node is received by a particular node with probability $q_1 = \frac{k_1}{n}$ $\left(q_2 = \frac{k_2}{n}\right)$ since the $X_i$ ($\tilde{X_i}$) are i.i.d.~for all receiver nodes. Noting the independence between the update generation and update transmission processes, at each cycle a particular node successfully receives a type I (type II) update with probability $p_1q_1$ ($p_2q_2$). Suppose a particular node has received a type I update packet during cycle $j$ and the next successful type I update packet delivery to that node is in cycle $j+M_1$. Then, this $M_1$ is a geometric random variable with success probability $p_1q_1$ and has moments
\begin{align}
E[M_1] &= \frac{1}{{p_1q_1}} = \frac{n}{p_1k_1} \label{moment1} \\  
E[M_1^2] &= \frac{2-{p_1q_1}}{p_1^2q_1^2} = \frac{2n^2}{p_1^2k_1^2}-\frac{n}{p_1k_1} \label{moment2}
\end{align}

\begin{figure}[t]
	\centering  \includegraphics[width=0.85\columnwidth]{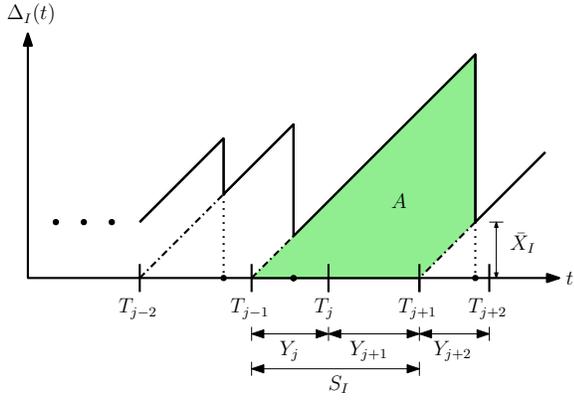}
	\caption{Sample age evolution $\Delta_{I}(t)$ of a node. Update cycle $j$ starts at time $T_{j-1}$ and lasts until $T_j$. Successful type I update deliveries are marked with $\bullet$ and in this figure, in cycles $j-1$, $j$ and $j+2$ a type I update is delivered successfully whereas in cycle $j+1$ no type I delivery occurred.}
	\label{fig:ageEvol}
	\vspace{-0.5cm}
\end{figure}

Note that in between update cycles $j$ and $j+M_1$ the source node may have sent type II updates. By using the success probability $p_1q_1$, we account for the time spent during type II update transmissions in between two successful type I update deliveries to this node. Fig.~\ref{fig:ageEvol} shows a sample $\Delta_I(t)$ evolution for a particular node. 
Correspondingly, if a certain node has received two successive type II updates in update cycles $j'$ and $j'+M_2$ then this $M_2$ is geometrically distributed with success probability $p_2q_2$ and its moments can be derived by changing the parameter from $p_1q_1$ to $p_2q_2$ in (\ref{moment1}) and (\ref{moment2}). We remark that $M_1$ and $M_2$ are independent from $Y_j$.

Noting the symmetry in the age of type I and type II streams at a particular receiver node, $\Delta_{I}$ and $\Delta_{II}$, respectively, here we derive $\Delta_{I}$ and deduce $\Delta_{II}$ from it by making necessary changes. In Fig.~\ref{fig:ageEvol}, $A$ denotes the shaded area and $S_I$ is its length. In other words, $S_I$ is the interarrival time of type I updates at a node. Inspecting Fig.~\ref{fig:ageEvol}, we find $E[A] = \frac{1}{2}E[S_I^2] + E[S_I]E[\bar{X}_I]$. Here, 
$\bar{X}_I$ denotes the transmission time of a successful type I update such that $E[\bar{X}_I] = E[X_i | i \in \mathcal{K}_I]$, where $\mathcal{K}_I$ is the set of earliest $k_1$ nodes that receive this type I update. Then, the average age of update stream I is given by
\begin{align}
\Delta_{I} =& \frac{E[A]}{E[S_I]} = E[\bar{X}_I] + \frac{E[S_I^2]}{2E[S_I]} \label{age_I_2}
\end{align}

Event $M_1 = m$ indicates two successive type I deliveries to a node in cycles $j$ and $j+m$ with $Y_j = X_{k_1:n}$ and $Y_{j+m} = X_{k_1:n}$ as well as $m-1$ consecutive type I failures in between. For these $m-1$ update cycles with no type I delivery we have
\begin{align}
(Y_j~|~\text{no type I delivery}) = \bar{Y}_j = \begin{cases} 
X_{k_1:n} & w.p.~\bar{p}_1\\
\tilde{X}_{k_2:n} & w.p.~\bar{p}_2 
\end{cases} \label{trans_time2}
\end{align}
with $\bar{p}_1 = \frac{p_1(1-q_1)}{1-p_1q_1}$ and $\bar{p}_2 = \frac{p_2}{1-p_1q_1}$ from Bayes' rule. Thus, $ S_I = X_{k_1:n} + \sum_{i=j+1}^{j+M_1-1} \bar{Y}_i$. For example, in Fig.~\ref{fig:ageEvol}, $M_1 = 2$ and therefore, we have $S_I = X_{k_1:n} + \bar{Y}$. Thus, we find
\begin{align}
	E[S_I] =& \hspace{1mm} E[X_{k_1:n}] + E[M_1-1]E[\bar{Y}] \label{ES} \\
	E[S_I^2] =& \hspace{1mm} E[X^2_{k_1:n}] + 2E[M_1-1]E[X_{k_1:n}]E[\bar{Y}] \nonumber \\ &+ E[M_1-1]Var[\bar{Y}]+E[(M_1-1)^2]E[\bar{Y}]^2 \label{ES^2}
\end{align}

In the following theorem, we determine the age of a type I update at an individual node using (\ref{age_I_2}).
\begin{theorem} \label{thm1}
Under the earliest $k_1$ and $k_2$ transmission scheme for type I and type II updates, respectively, the average type I age at an individual node is
\begin{align}
\Delta_{I} =& \frac{1}{k_1} \sum_{i=1}^{k_1} E[X_{i:n}] + \frac{p_1E[X^2_{k_1:n}] + p_2E[\tilde{X}^2_{k_2:n}] }{2p_1E[X_{k_1:n}]+2p_2E[\tilde{X}_{k_2:n}]} \nonumber \\ &+ \frac{p_2^2nE[\tilde{X}_{k_2:n}]^2 + p_1p_2(2n-k_1)E[X_{k_1:n}] E[\tilde{X}_{k_2:n}]}{p_1k_1(p_1E[X_{k_1:n}]+p_2E[\tilde{X}_{k_2:n}])} \nonumber \\ &+ \frac{p_1^2(n-k_1)E[X_{k_1:n}]^2}{p_1k_1(p_1E[X_{k_1:n}]+p_2E[\tilde{X}_{k_2:n}])}\label{thm1_res}
\end{align}
\end{theorem} 
\begin{Proof}
The first term comes from $E[\bar{X}_I]$ as
\begin{align}
E[\bar{X}_I]\!=&E[X_j| j \in \mathcal{K}_I]\!=\sum_{i=1}^{k_1} E[X_{i:n}] Pr[j=i| j \in \mathcal{K}_I ] \label{trans_success}
\end{align}	
where $Pr[j=i| j \in \mathcal{K}_I] = \frac{1}{k_1}$ since we have $k_1$ out of $n$ nodes selected independently and identically in $\mathcal{K}_I$. The claim follows by substituting (\ref{ES}), (\ref{ES^2}) and (\ref{trans_success}) back into (\ref{age_I_2}), and replacing $E[M_1]$ and $E[M_1^2]$ by (\ref{moment1}) and (\ref{moment2}). Moments of $\bar{Y}$ follow from (\ref{trans_time2}).
\end{Proof}

When $p_1=1$, i.e., the source node generates only type I updates, $\Delta_{I}$ given in (\ref{thm1_res}) reduces to the single stream result in \cite[Theorem 2]{Zhong17a}. From the symmetry of the network model, the average age expression of type II update stream at an individual node, $\Delta_{II}$, can be found upon defining $\bar{Y}$ accordingly and replacing $k_1$, $p_1$ and $X_{k_1:n}$ with $k_2$, $p_2$ and $\tilde{X}_{k_2:n}$ in (\ref{thm1_res}). When the service times of the packets of the same kind are i.i.d.~shifted exponential random variables and $n$ is large, we can further simplify (\ref{thm1_res}) as follows.

\begin{corollary} \label{corr1}
	For large $n$ and $n>k_i$ we set $k_i = \alpha_in$ for $i=1, 2$. For shifted exponential transmission times $X$ and $\tilde{X}$ with parameters $(\lambda,c)$ and $(\tilde{\lambda}, \tilde{c})$ for type I and type II updates, respectively, $\Delta_{I}$ can be approximated as  
	\begin{align}
		\Delta_I \approx&\hspace{1mm} c+ \frac{1}{\lambda} + \frac{1-\alpha_1}{\lambda \alpha_1} \log(1-\alpha_1) \nonumber \\ &+ \frac{(2-\alpha_1)p_1^2 \delta^2_1(\alpha_1) + 2p_1p_2(2-\alpha_1)\delta_1(\alpha_1)\delta_2(\alpha_2)}{2p_1\alpha_1(p_1\delta_1(\alpha_1)+p_2\delta_2(\alpha_2))} \nonumber \\
		&+ \frac{p_2(p_1\alpha_1+2p_2)\delta^2_2(\alpha_2)}{2p_1\alpha_1(p_1\delta_1(\alpha_1)+p_2\delta_2(\alpha_2))}	\label{corr1_res}
	\end{align}
	where we denote
	\begin{align}
	\delta_1(\alpha_1) = c\!-\!\frac{\log(1\!-\!\alpha_1)}{\lambda}, \quad 
	\delta_2(\alpha_2) = \tilde{c}\!-\!\frac{\log(1\!-\!\alpha_2)}{\tilde{\lambda}} \label{deltas}
	\end{align}
\end{corollary}
\begin{Proof}
	For the first term in (\ref{thm1_res}) we have 
	\begin{align}
	\frac{1}{k_1} \sum_{i=1}^{k_1} E[X_{i:n}] &=  c + \frac{H_n}{\lambda} - \frac{1}{k_1\lambda}\sum_{i=1}^{k_1} H_{n-i} \label{delta1_exp}\\ &\approx c+ \frac{1}{\lambda} + \frac{1-\alpha_1}{\alpha_1\lambda }\log(1-\alpha_1) \label{delta1_exp2}
	\end{align}
	where (\ref{delta1_exp}) is obtained using (\ref{ord1})-(\ref{ord3}). We have $\sum_{i=1}^{k_1} H_{n-i} = \sum_{i=1}^{n-1} H_i - \sum_{i=1}^{n-k_1-1} H_i$ and the series identity $\sum_{i=1}^{k_1} H_i  = (k_1+1)(H_{k_1+1}-1)$. Using these and the fact that for large $n$ $H_i \approx \log(i) + \gamma$ yields (\ref{delta1_exp2}). Note that
	\begin{align}
	E[X_{k_1:n}] \!&=\! c\!+\! \frac{H_n \!-\! H_{n-k_1}}{\lambda} \!\approx\! c\!-\!\frac{\log(1\!-\!\alpha_1)}{\lambda} \!=\! \delta_1(\alpha_1)\\
	E[\tilde{X}_{k_2:n}] &\!=\!\tilde{c}\!+\!\frac{H_n\!-\!H_{n-k_2}}{\tilde{\lambda}}\!\approx\! \tilde{c}\!-\!\frac{\log(1\!-\!\alpha_2)}{\tilde{\lambda}}\!=\!\delta_2(\alpha_2)
	\end{align}
	
	Lastly, we note that $E[X^2_{k_1:n}] \approx (E[X_{k_1:n}])^2$ for large $n$. This is because the sequence $G_{n}$ converges to $\frac{\pi^2}{6}$ as $n$ goes to $\infty$. Thus, as $n$ increases $G_{n-k_1} = G_{(1-\alpha_1)n}$ also goes to $\frac{\pi^2}{6}$. With this, $G_n - G_{n-k_1}$ term in (\ref{ord3}) approaches $0$ yielding the claim. Likewise, we have $E[\tilde{X}^2_{k_2:n}] \approx (E[\tilde{X}_{k_2:n}])^2$. Combining all these to calculate the moments of $\bar{Y}$ and taking $k_i = \alpha_in$ for $i=1, 2$ gives (\ref{corr1_res}).
	\end{Proof}

We remark that when $p_1 =1$, i.e., the source node only generates type I updates, Corollary~\ref{corr1} reduces to the single update stream result in \cite[Corollary 2]{Zhong17a}. From the symmetry of the system, average age of type II updates at an individual node, $\Delta_{II}$, can be also approximated as in Corollary~\ref{corr1}. 

We note that under the earliest $k_1$ and $k_2$ transmission scheme for updates of type I and type II, respectively, the average age is a function of ratios $\alpha_1$ and $\alpha_2$ for large $n$. Thus, the average age still converges to a constant even though the multicast network is shared across two update streams. 

We can minimize $\Delta_{I}$ and $\Delta_{II}$ by selecting the stopping thresholds $k_1, k_2$. Since we have two age processes to minimize, we can consider the following optimization problem
\begin{align}
\label{problem1}
\min_{\{k_{1}, k_{2} \}}  \quad & \beta \Delta_I + (1-\beta) \Delta_{II} \nonumber \\
\mbox{s.t.} \quad & 0 \leq \beta \leq 1 \nonumber \\
\quad & k_i\in \{1, \dots, n \}, \quad i=1, 2
\end{align}

Thus, by varying $\beta$ we can assign weights to the update types I and II during optimization. While extreme cases of $\beta = 1$ and $\beta = 0$ assign absolute priority to update type I and type II, respectively, every other $\beta$ value weighs the update streams in between accordingly. Equivalently, we can solve (\ref{problem1}) over $\alpha_1$ and $\alpha_2$. In that case, instead of the last constraint in (\ref{problem1}) we have $ 0 < \alpha_i < 1$, $i=1, 2$.

\begin{lemma} \label{lemma1}
	When $n$ is large, for any given $k_1 \in \{1, \dots, n\}$ to achieve the best type I average age performance, i.e., to minimize $\Delta_{I}$, it is optimal to select $k^*_2 = 1$.
\end{lemma}
\begin{Proof}
	This is the $\beta = 1$ case. We observe that for a given $k_1$ the three terms in (\ref{corr1_res}) and $\delta_1 = E[X_{k_1:n}]$ become constant. The only parameter we select to optimize $\Delta_{I}$ is $k_2$ and for large $n$ it only appears in (\ref{corr1_res}) as $\delta_2(k_2) = E[\tilde{X}_{k_2:n}]$. $\Delta_{I}$ given in (\ref{corr1_res}) is in the following form
	\begin{align}
		\Delta_{I} = c_1 + \frac{c_2 \delta^2_1 + c_3 \delta_1 \delta_2(k_2) + c_4 \delta^2_2(k_2)}{c_5 \delta_1 + c_6 \delta_2(k_2)} \label{age_const}
	\end{align}
	where $c_1, \dots, c_6$ and $\delta_1$ are positive numbers. Noting that $\delta_2(k_2)$ is also positive, (\ref{age_const}) is an increasing function of $\delta_2(k_2)$ when $c_2c_6 < c_3c_5$ condition is satisfied. From (\ref{corr1_res}), we see that this condition is met. Thus, to minimize (\ref{age_const}) we need to select the smallest possible $\delta_2(k_2)$ which is given by the smallest possible $k_2$ since $\delta_2(k_2)$ is monotonically increasing in $k_2$. Thus, $k^*_2 = 1$ minimizes the type I average age, $\Delta_{I}$, for any given $k_1$.	
\end{Proof}

\begin{figure}[t]
	\begin{center}
		\subfigure[\label{fig:pareto}]{%
			\includegraphics[width=0.52\linewidth]{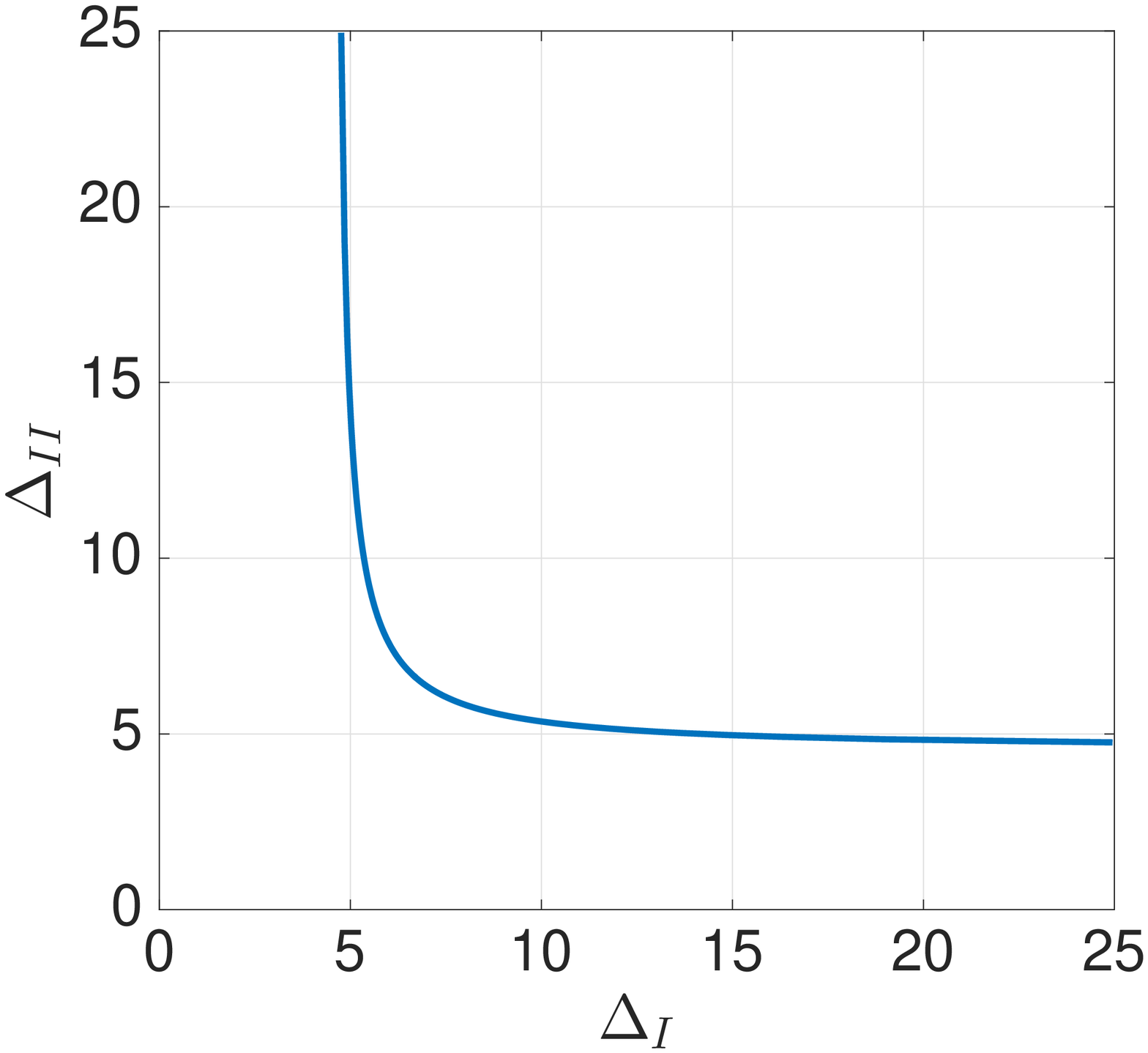}}
			\hspace{-5.75mm}
		\subfigure[\label{fig:pareto_exo}]{%
			\includegraphics[width=0.52\linewidth]{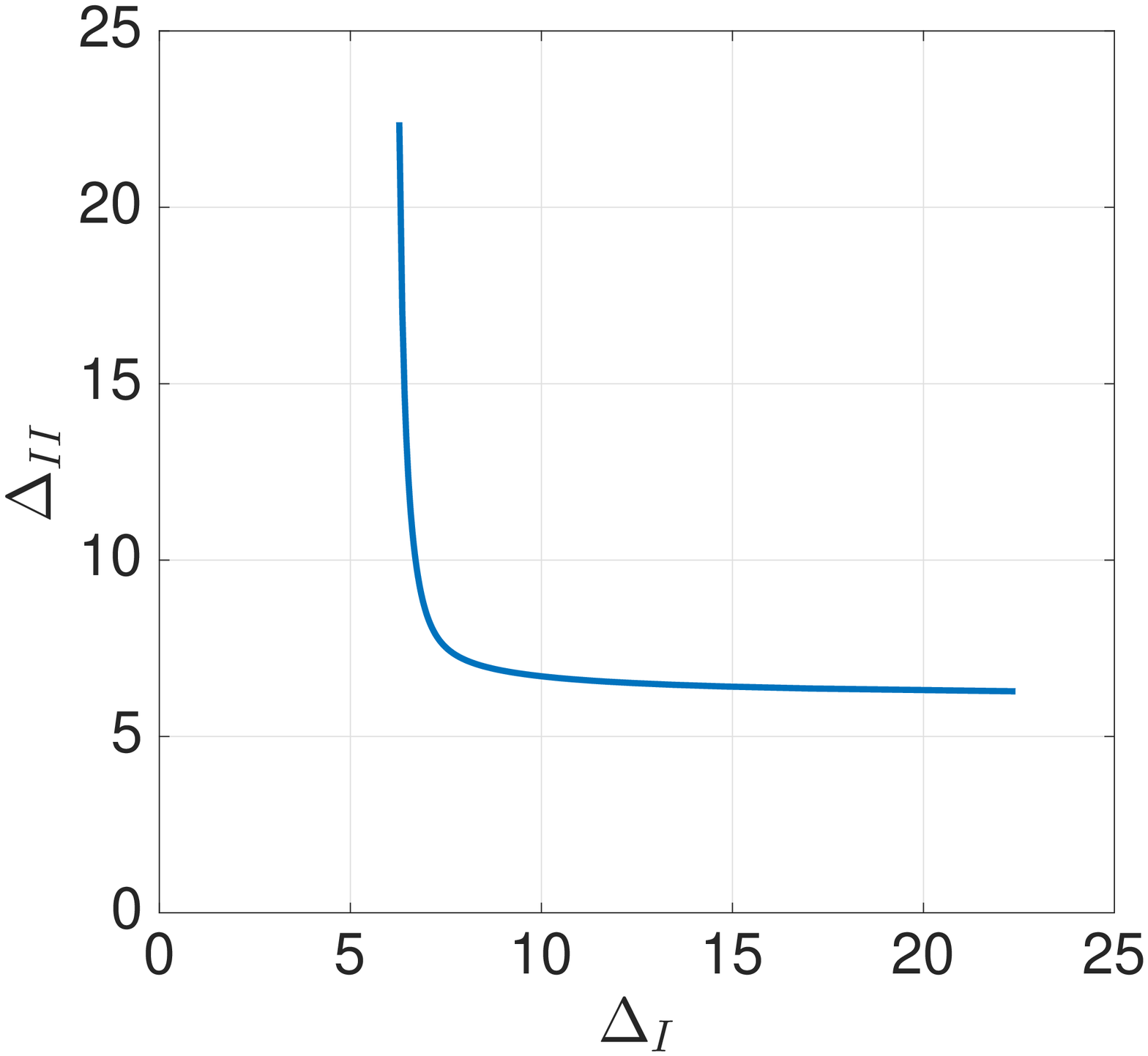}}
			\hspace{-5.75mm}
		\caption{Pareto optimal curve for jointly minimizing $\Delta_{I}$ and $\Delta_{II}$ with $0< \beta < 1$, $p_1 = p_2 = 0.5$ and $(\lambda,c) = (\tilde{\lambda}, \tilde{c}) = (1,1)$ when (a) updates are generated at-will, and (b) when updates arrive exogenously. }
		\label{Fig:Add_sim_result}
	\end{center}
	\vspace{-5mm}
\end{figure}

Similarly, when we are only interested in minimizing the type II average age, $\Delta_{II}$, i.e., $\beta = 0$, it is optimal to select $k^*_1 =1$. Fig.~\ref{fig:pareto} shows the pareto optimal curve obtained upon numerically solving (\ref{problem1}) for $p_1 = p_2 = 0.5$ and shifted exponential link delays with $(1,1)$ for both types of updates.

\subsection{Exogenous Update Arrivals}
Updates arrive at the source node exogenously as a Poisson process with a total rate of $\mu$. Each arriving update is of type I with probability $p_1$ and of type II with probability $p_2$ where $p_1+p_2 = 1$. Thus, type I update stream arrives as a Poisson process with rate $\mu_1=\mu p_1$ and analogously, type II update stream arrives as a Poisson process with rate $\mu_2=\mu p_2$. The source node implements the earliest $k_1$, $k_2$ transmission scheme for type I and type II update packets, respectively. When there is a packet in service, the source node discards any other arriving update packet. Thus, update types do not have priority over each other during transmission. However, during joint optimization of the average age of both streams, $\Delta_{I}$ and $\Delta_{II}$, by varying $\beta$ we can prioritize them. 

When an update is completed, i.e., transmitted to the earliest $k_i$ nodes, $i=1,2$ depending on the type of the update packet, the system stays idle until the next update packet of any kind arrives. We denote this idle period with random variable $Z$ which is an exponential random variable with rate $\mu p_1 + \mu p_2 = \mu$. Time spent during transmission of an update is the busy period $Y_B$ which is equivalent to (\ref{trans_time})
since after each idle period with probability $p_1$ a type I update and with probability $p_2$ a type II update goes into service. Thus, an update cycle, $Y$, is equivalent to $Y_B+Z$. Note that $Z$ does not depend on $Y_B$ since it is memoryless. We have random variables $M_1$ and $M_2$ which are geometrically distributed with $p_1q_1$ and $p_2q_2$, respectively, as in Section~\ref{atwill}. Also note that $M_1$ and $M_2$ are independent from update cycle $Y$. 

Age of type I updates at an individual node is given by (\ref{age_I_2}). However, type I update interarrival to a node, $S_I$, is now equal to $S_I = X_{k_1:n} + Z + \sum_{i=j+1}^{j+M_1-1} \bar{Y}_i = X_{k_1:n} + \sum_{i=1}^{M_1-1} (\bar{Y}_B)_i + \sum_{i=1}^{M_1}Z_i$ where $\bar{Y}_B$ is equivalent to (\ref{trans_time2}). Then,
\begin{align}
	E[S_I] =& \hspace{1mm} E[X_{k_1:n}] + E[M_1-1]E[\bar{Y}_B] + E[M_1]E[Z] \label{ES_2} \\
	E[S_I^2] =& \hspace{1mm} E[X^2_{k_1:n}] + 2E[M_1-1]E[X_{k_1:n}]E[\bar{Y}_B] \nonumber \\ &+ 2E[M_1]E[X_{k_1:n}]E[Z] + E[M_1]Var[Z]  \nonumber \\ &+ E[M_1-1]Var[\bar{Y}_B]+E[(M_1-1)^2]E[\bar{Y}_B]^2 \nonumber \\ &+ 2E[M^2_1-M_1]E[\bar{Y}_B]E[Z] +E[M^2_1]E[Z]^2  \label{ES^2_2}
\end{align}

In the following theorem, we determine the age of a type I update at an individual node when the update streams arrive exogenously to the source node.

\begin{theorem} \label{thm2}
Under the earliest $k_1$ and $k_2$ transmission scheme for type I and type II updates that arrive to the source node as Poisson processes with rates $\mu_1$ and $\mu_2$, respectively, the average type I age at an individual node is
\begin{align}
\Delta_{I} =& \frac{1}{k_1} \sum_{i=1}^{k_1} E[X_{i:n}] + \frac{E[S^2_I]}{2E[S_I]} \label{thm2_res}
\end{align}
where first and second moments of $S_I$ are as in (\ref{ES_2}) and (\ref{ES^2_2}).
\end{theorem}

The proof of Theorem~\ref{thm2} follows accordingly from that of Theorem~\ref{thm1}. Note that when $p_1=1$, (\ref{thm2_res}) reduces to the building block result in \cite[Theorem 1]{Buyukates18b}. By making the corresponding replacements as in Section~\ref{atwill} we can obtain the average age expression of type II update stream, $\Delta_{II}$. When the service times of the packets of the same kind are i.i.d.~shifted exponential random variables and $n$ is large, we can further simplify (\ref{thm2_res}) as follows.
\begin{corollary} \label{corr2}
	For large $n$ and $n>k_i$ we set $k_i = \alpha_in$ for $i=1, 2$. For shifted exponential transmission times $X$ and $\tilde{X}$ with parameters $(\lambda,c)$ and $(\tilde{\lambda}, \tilde{c})$ for type I and type II updates, respectively, $\Delta_{I}$ can be approximated as  
\begin{align}
\Delta_I \approx& \hspace{1mm} c+ \frac{1}{\lambda} + \frac{1-\alpha_1}{\lambda \alpha_1} \log(1-\alpha_1) \nonumber \\ &+ \frac{\mu p_1^2 (2-\alpha_1) \delta^2_1(\alpha_1) + 2\mu p_1p_2(2-\alpha_1)\delta_1(\alpha_1)\delta_2(\alpha_2) }{2p_1\alpha_1(\mu p_1 \delta_1(\alpha_1) + \mu p_2 \delta_2(\alpha_2) +1)} \nonumber \\ &+ \frac{ \mu p_2(2p_2+p_1\alpha_1)\delta^2_2(\alpha_2)}{2p_1\alpha_1(\mu p_1 \delta_1(\alpha_1) + \mu p_2 \delta_2(\alpha_2) +1)} \nonumber \\ &+ \frac{2\mu p_2\delta_2(\alpha_2) + \mu p_1(2-\alpha_1)\delta_1(\alpha_1) +1 }{\mu p_1\alpha_1(\mu p_1 \delta_1(\alpha_1) + \mu p_2 \delta_2(\alpha_2) +1)}
\end{align}
where $\delta_1(\alpha_1)$ and $\delta_2(\alpha_2)$ are as in (\ref{deltas}).
\end{corollary}

The proof of Corollary~\ref{corr2} follows accordingly from that of Corollary~\ref{corr1}. We note that when $p_1=1$, Corollary~\ref{corr2} reduces to the result in \cite[Corollary 2]{Buyukates18b}. The corresponding approximate expression for $\Delta_{II}$ can also be derived in the same manner. We also observe that when the earliest $k_1$, $k_2$ transmission scheme is utilized for type I and type II updates, respectively, the average age of either update type is also a function of ratios $\alpha_1$ and $\alpha_2$. Thus, although multiple update streams that arrive exogenously use the same network, an average age that does not depend on $n$ can be achieved for either update stream.

When we minimize $\beta\Delta_{I}+(1-\beta)\Delta_{II}$ with $\beta \in (0,1)$ by selecting $k_1, k_2 \in \{1,\dots, n \}$ as in (\ref{problem1}), we obtain the pareto optimal curve as shown in Fig.~\ref{fig:pareto_exo}. For the cases of $\beta=1$ and $\beta=0$ which correspond to the individual optimization of $\Delta_{I}$ and $\Delta_{II}$, respectively, we observe that when $\beta=1$ it is optimal to select $k^*_2=1$ to obtain the minimum $\Delta_{I}$, and when $\beta=0$ selecting $k^*_1=1$ gives the minimum $\Delta_{II}$. The proof of this claim is similar to that of Lemma~\ref{lemma1} since random variable $Z$ is also positive and independent of $k_1$ and $k_2$.

\bibliographystyle{unsrt}
\bibliography{IEEEabrv,lib}

\end{document}